\documentclass[%
 reprint,
superscriptaddress,
nofootinbib,
 amsmath,amssymb,
 aps,
]{revtex4-1}

\usepackage{braket}
\usepackage{soul}
\usepackage{multirow}
\usepackage{graphicx}% Include figure files
\usepackage{dcolumn}% Align table columns on decimal point
\usepackage{bm}% bold math
\usepackage[usenames, dvipsnames]{xcolor}
\usepackage[colorlinks=true, urlcolor=MidnightBlue, linkcolor=MidnightBlue, citecolor=Mahogany, pdfborder={0 0 0}]{hyperref}% add hypertext capabilities
\usepackage[capitalise]{cleveref}

\begin{document}

%\preprint{APS/123-QED}

\title{The interplay of transverse degrees of freedom and axial-vector mesons with short-distance constraints in \texorpdfstring{$g-2$}{g-2}}

\author{Pere Masjuan}\email{masjuan@ifae.es}
\affiliation{
Grup de F{\'i}sica Te{\`o}rica, Departament de F{\'i}sica, Universitat Aut{\`o}noma de Barcelona}
\affiliation{
Institut de F{\'i}sica d'Altes Energies (IFAE) and \\
The Barcelona Institute of Science and Technology, \\
Universitat Aut{\'o}noma de Barcelona, E-08193 Bellaterra (Barcelona), Spain
}
\author{Pablo Roig}\email{proig@fis.cinvestav.mx}
\affiliation{
Centro de Investigaci{\'o}n y de Estudios Avanzados del IPN (Cinvestav), \\
Apdo.~Postal~14-740, 07000 Ciudad de M{\'e}xico, M{\'e}xico
}
\author{Pablo Sanchez-Puertas}\email{psanchez@ifae.es}
\affiliation{
Institut de F{\'i}sica d'Altes Energies (IFAE) and \\
The Barcelona Institute of Science and Technology, \\
Universitat Aut{\'o}noma de Barcelona, E-08193 Bellaterra (Barcelona), Spain
}

\begin{abstract}
We revisit well-known short-distance constraints relating the hadronic light-by light Green's function to the $\langle VVA \rangle$ one, that have been a subject of debate over the past years in the context of the muon $(g-2)$. Specifically, we identify a relation among the longitudinal and transverse degrees of freedom that is enforced by the axial anomaly that, by contrast, has not received attention in the past. Such relation allows, among other things, to overcome the problem of basis ambiguities when describing axial-vector mesons transition form factors, but further applications are discussed as well, with special focus on the role of axial-vector mesons in the HLbL contribution to the muon $(g-2)$. Our results should also contribute to a better understanding of the, so far, controversial interplay among short-distance constraints with longitudinal and transverse  degrees of freedom, such as axial-vector mesons. This is key to confront the theoretical and experimental result for the muon $(g-2)$ that, currently, exhibits a $4.2\sigma$ tension.
\end{abstract}

\maketitle

\section{\label{sec:intro}Introduction}

The anomalous magnetic moment of the muon, $a_{\mu}\equiv(g_{\mu}-2)/2$~\cite{Jegerlehner:2009ry, Aoyama:2020ynm} is an extremely interesting observable. Over time, it helped consolidating our understanding of the Standard Model (SM) of particle physics at the multi-loop level, thanks to the increased precision reached by the experiment and the successive refinements in its computation. Actually, its most recent measurements~\cite{Bennett:2002jb,Bennett:2004pv,Bennett:2006fi,Abi:2021gix} have achieved enough accuracy to contribute prominently in the search for new physics  in the intensity frontier~\cite{Lindner:2016bgg,Crivellin:2018qmi,Keshavarzi:2019abf}, and to constraint tightly new-physics scenarios.  Particularly interesting is the current $4.2\sigma$ discrepancy with respect to the SM prediction~\cite{Jegerlehner:2017gek,Keshavarzi:2018mgv, Davier:2019can, Aoyama:2020ynm}, that also benefits from the extreme precision of QED \cite{Aoyama:2017uqe, Aoyama:2012wk, Aoyama:2014sxa}, EW \cite{Gnendiger:2013pva, Czarnecki:2002nt, Knecht:2002hr} and subleading hadronic corrections \cite{Kurz:2014wya, Colangelo:2014qya} (with a smaller $\sim2.2\sigma$ discrepancy if $\tau$ data is used instead \cite{Davier:2013sfa, Miranda:2020wdg}). This long-lasting puzzle triggered two new different experiments aiming to consolidate and to improve the  experimental accuracy: one at Fermilab~\cite{Grange:2015fou}, whose first  results were %just
published \cite{Abi:2021gix}, and one at J-PARC~\cite{Abe:2019thb}, that together will unveil the nature of the current anomaly at a new level of precision.

This effort must be accompanied by   a precise theoretical prediction for $a_{\mu}$ with similar accuracy as its experimental counterpart. Its bottleneck is the hadronic uncertainties, dominated by the hadronic vacuum polarization (HVP) and the hadronic light-by-light (HLbL) contributions. For that reason, the last years have witnessed a collective effort in the theory community to improve previous estimates for both hadronic contributions. With the increased accuracy that has been achieved for the HVP~\cite{Jegerlehner:2017gek,Keshavarzi:2018mgv, Davier:2019can} and foreseen improvements in
$e^+e^-\to\textrm{hadrons}$ cross section measurements (see also the alternative approach from MUonE Coll.~\cite{Banerjee:2020tdt} and advances from lattice QCD~\cite{DellaMorte:2017dyu, Blum:2018mom, Giusti:2019xct, Davies:2019efs, Gerardin:2019rua, Borsanyi:2020mff}), it is mandatory to ameliorate the precision of previous estimates for the HLbL contribution (see e.g., Refs.~\cite{Knecht:2001qf,Hayakawa:2001bb,Bijnens:2001cq,Melnikov:2003xd, Jegerlehner:2005fs,Prades:2009tw,Masjuan:2012wy,Escribano:2013kba,Masjuan:2014rea,Roig:2014uja, Benayoun:2014tra}). Such an effort has been undertaken by several groups~\cite{Pascalutsa:2012pr, Masjuan:2012qn, Pauk:2014rta, Jegerlehner:2017gek, Colangelo:2014dfa, Colangelo:2014pva, Escribano:2015nra,Colangelo:2015ama, Masjuan:2015lca, Escribano:2015yup, Masjuan:2015cjl,  Bijnens:2016hgx, Masjuan:2017tvw, Colangelo:2017qdm, Colangelo:2017fiz,  Guevara:2018rhj, Hoferichter:2018dmo, Knecht:2018sci, Hoferichter:2018kwz, Eichmann:2019tjk,  Raya:2019dnh,  Eichmann:2019bqf, Leutgeb:2019gbz, Cappiello:2019hwh} and the situation is encouraging, with additional promising results from lattice QCD as well~\cite{Blum:2014oka, Green:2015sra, Blum:2015gfa, Meyer:2018til,Chao:2021tvp}. 

From a theoretical point of view, the HLbL has always been approached via its splitting in terms of high and low energies, with the latter  understood as a sum of individual contributions. In particular, those  that have been thought to play the dominant role  at these low energies, \textit{the pseudoscalar poles}, have been computed to an accuracy that is enough for the upcoming experiments~\cite{Masjuan:2017tvw, Hoferichter:2018dmo, Hoferichter:2018kwz, Gerardin:2019vio, Raya:2019dnh}. In addition, subleading contributions such as $\pi^+\pi^-$ exchange have been calculated to an extraordinary precision as well \cite{Colangelo:2017qdm, Colangelo:2017fiz}. 

Still, there are a few additional contributions for which the present level of accuracy needs to be improved.  
Among them, one finds the axial-vector-mesons contributions \cite{Pauk:2014rta,Jegerlehner:2017gek, Roig:2019reh}, bringing up two different problems. On the one hand, there is the lack of precise-enough experimental data for the relevant transition form factors (TFFs) entering the calculation \cite{Roig:2019reh,Szczurek:2020hpc, Zanke:2021wiq, Lebiedowicz:2021gub}. On the other hand, Ref.~\cite{Roig:2019reh} put forward the problematic related to the choice of a tensor basis to describe such TFFs. Basis choice is irrelevant for on-shell processes, which is no more true when going off-shell, as it is the case in the HLbL.  This leads to potential ambiguities that require a better understanding in order to avoid or gauge systematic errors.

Furthermore, beyond individual meson contributions, we still lack a sensible estimate for the whole high-energy part of the HLbL, that cannot be described with a finite number of meson contributions~\cite{Masjuan:2007ay}. In particular, we shall be concerned with the high-energy regime first derived by Melnikov and Vainshtein (MV)~\cite{Melnikov:2003xd} on the basis of the operator product expansion (OPE), and that allows to connect the HLbL to the $\langle VVA \rangle$ (vector-vector-axial three-point) Green’s function. Based on that connection, MV showed that pseudoscalar-pole contributions could not accommodate well-established properties of the $\langle VVA \rangle$ Green's function, advocating for an alternative model that, however, is at odds with the analytic properties of the HLbL (see for instance \cite{Jegerlehner:2009ry}).
To improve on that situation, alternative approaches have been proposed~\cite{Nyffeler:2009tw,Colangelo:2019lpu,Colangelo:2019uex} that, in contrast with \cite{Melnikov:2003xd}, focused exclusively on purely longitudinal contributions, obtaining notably different results. 
Unfortunately, such models leave aside the role of transverse contributions, while the OPE constraints shall be imposed for the full HLBL tensor. Interestingly enough, we shall show that this situation is deeply interrelated to the aforementioned problematic related to the tensor basis of the axial-vector mesons TFFs, and is partly responsible for the seemingly different results obtained so far in the literature for the axial-vector meson contributions to $a_{\mu}$. All in all, it is fair to say that potentially large systematic errors may obscure current estimates for axial mesons, situation requiring a better understanding.

In this work, we revisit the short-distance constraints first derived in \cite{Melnikov:2003xd}, carefully examining the role of the lightest pseudoscalar mesons, as well as their interplay with transverse contributions. As a result, we find a nontrivial relation among them, that is dictated by the axial anomaly in the chiral limit and that has not received attention in the past, but only recently in  Ref.~\cite{Knecht:2020xyr}. Remarkably, such connection allows to overcome the basis problem for axial-vector mesons once the relevant degrees of freedom in the model are specified, which is a novel result. 
As such, it allows for a better understanding of present estimates of axial-vector mesons to $a_{\mu}$, as well as their non-pole contribution. Last, but not least, it sheds light on current approaches to estimate the high-energy regime discussed in \cite{Melnikov:2003xd}, and suggests possible lines for improvement concerning the role of heavy pseudoscalars beyond the chiral limit.

The article is organized as follows. \cref{sec:main} describes the high-energy limit of interest, that connects the HLbL and the $\langle VVA \rangle$ Green's function. In particular, it points to the important role that transverse degrees of freedom play in fulfilling the anomaly. This has important  consequences that we discuss in the subsequent sections. To start with, \cref{sec:ppoles} considers the pseudoscalar-pole contribution to the HLbL, including the difficulties to fulfill the anomaly with pseudoscalar poles alone. Then, \cref{sec:axial} illustrates the interplay of transverse and longitudinal degrees of freedom in a model of axial and pseudoscalar mesons. Finally, \cref{sec:simplemodel} illustrates our main points within a simple but efficient Regge model re-summing the infinite tower of axial-vector mesons. It also includes a parallel discussion using recent holographic models.

\section{Main definitions\label{sec:main}}

The relevant object for computing the HLbL contribution to the anomalous magnetic moment of the muon, $a_{\mu}^{\textrm{HLbL}}$, is the HLbL Green's function, defined as~\cite{Masjuan:2017tvw,Knecht:2018sci,Roig:2019reh}
\begin{multline}\label{eq1}
    \Pi^{\mu\nu\lambda\sigma}(q_1,q_2,q_3) = 
    \int d^4xd^4yd^4z e^{i(q_1\cdot x +q_2\cdot y +q_3\cdot z)} \\ 
    \times \bra{0} T\{j^{\mu}(x) j^{\nu}(y) j^{\lambda}(z) j^{\sigma}(0) \} \ket{0},
\end{multline}
with $j^{\mu} = \bar{q}\gamma^{\mu} \mathcal{Q}  q$, $\mathcal{Q}=\operatorname{diag}(2/3,-1/3,-1/3)$ the charge operator and $q=(u,d,s)^T$. This defines $q_{1,2,3}$ momenta outgoing, while it is customary to take the combination $q_4=q_1+q_2+q_3$ as incoming momentum. With these conventions, its contribution to $a_{\mu}$ can be readily evaluated using the method outlined in Ref.~\cite{Roig:2019reh}.

In particular, we shall focus here on the high-energy behavior  derived in Ref.~\cite{Melnikov:2003xd} (see also \cite{Colangelo:2019uex,Ludtke:2020moa}), that allows to solve ambiguities arising when including axial-vector-mesons contributions to HLbL, deeply interrelated with the pseudoscalar poles, a point often missed. Such limit concerns the \textit{mixed} region in which $q_1^2\sim q_2^2 \gg q_{12,3,4}^2$, with $q_{12}\equiv q_1 +q_2$, and can be obtained on the basis of the OPE (see also \cite{Bijnens:2019ghy,Bijnens:2020xnl,Bijnens:2021jqo} for the three hard-photon case). Introducing $\hat{q} \equiv (q_1 -q_2)/2 \equiv \bar{q}_{12}/2$, the OPE result for the time-ordered product of two vector currents reads, for $\hat{q}^2<0$ ~\cite{Melnikov:2003xd},
\begin{multline}\label{eq:opegen}
    \int d^4x d^4y e^{i(q_1\cdot x +q_2\cdot y)}  T\{ j^{\mu}(x) j^{\nu}(y)\}   = \\ \frac{-2}{\hat{q}^2} \epsilon^{\mu\nu\alpha \hat{q}} \int d^4z e^{i q_{12}\cdot z}  j_{5\alpha}(z)  +\mathcal{O}\left(\frac{\Lambda_{\textrm{QCD}}^{2}}{\hat{q}^{2}}\right),
\end{multline}
with $j_5^{\mu} = \bar{q}\gamma^{\mu}\gamma^5 \mathcal{Q}^2 q$, $\epsilon^{\mu\nu\rho q_i}\equiv \epsilon^{\mu\nu\rho \alpha}q_{i\alpha}$, and $\epsilon^{0123}=1$, one can infer the corresponding short-distance QCD behavior for the HLbL tensor defined in Eq.~(\ref{eq1})
\begin{multline}\label{eq:OPE1}
    \Pi^{\mu\nu\lambda\sigma}_{\textrm{OPE}}(q_1,q_2,q_3) = \frac{2i\epsilon^{\mu\nu\alpha \hat{q}}}{\hat{q}^2} \times i\int d^4z d^4w e^{i(q_{12}\cdot w +q_3\cdot z)} \\   \times \bra{0} T\{ j_{5\alpha}(w) j^{\lambda}(z) j^{\sigma}(0)   \} \ket{0}  + \mathcal{O}\left(\frac{\Lambda_{\textrm{QCD}}^{2}}{\hat{q}^{2}}\right),
\end{multline}
where $\Lambda_{\textrm{QCD}}$, of the order of few hundred MeV, suppresses the neglected contributions that we do not show in the following. 
This can be further expressed in terms of the $\langle VVA \rangle$ Green's function defined as~\cite{Knecht:2003xy,Jegerlehner:2005fs}
\begin{multline}\label{eq:VVAGF}
     i\int d^4x d^4y e^{i(q_1\cdot x +q_2\cdot y)} \bra{0} T\{  V_{\mu}^{a}(x) V_{\nu}^{b}(y) A_{\rho}^{c}(0)  \} \ket{0} \\ 
     \equiv \mathcal{W}_{\mu\nu\rho}^{abc}(q_1,q_2) = \mathcal{W}_{\mu\nu\rho}(q_1,q_2) \operatorname{tr}(t^c \{ t^a,t^b\}),
\end{multline}
where $V_{\mu}^a = \bar{q}\gamma_{\mu}t^a q$, $A_{\mu}^a = \bar{q}\gamma_{\mu}\gamma^5t^a q$ and $t^a = \lambda^a/2$, with $\lambda^a$ the Gell-Mann matrices, allowing to recast \cref{eq:OPE1} as
\begin{equation}\label{eq:hlblope}
    \Pi^{\mu\nu\lambda\sigma}_{\textrm{OPE}}(q_1,q_2,q_3) = 
    \frac{4i}{\hat{q}^2} \operatorname{tr}(\mathcal{Q}^4) \epsilon^{\mu\nu\alpha \hat{q}} \mathcal{W}^{\lambda\sigma}_{\ \ \ \alpha}(q_3,-q_4),
\end{equation}
that we shall refer in the following as the ``OPE constraint''. Alternatively, performing the axial current isospin decomposition either in the singlet-octet or quark-flavor basis ($j_{5\mu} = \sum c_a A_{\mu}^a$ with $c_a = \operatorname{tr}[\lambda^a \mathcal{Q}^2]$), the equation above can be expressed as (we omit here momentum arguments)
\begin{align}\label{eq:opeisospin}
\Pi^{\mu\nu\lambda\sigma}_{\textrm{OPE}} &{}=
    \frac{2i \epsilon^{\mu\nu\alpha \hat{q}}}{9\hat{q}^2}\left( \mathcal{W}_{(3)\alpha}^{\lambda\sigma}  +\frac{1}{3}\mathcal{W}_{(8)\alpha}^{\lambda\sigma}  +\frac{8}{3}\mathcal{W}_{(0)\alpha}^{\lambda\sigma} \right) \nonumber \\
    &{}= \frac{2i \epsilon^{\mu\nu\alpha \hat{q}}}{9\hat{q}^2}\left( \mathcal{W}_{(3)\alpha}^{\lambda\sigma} +\frac{25}{9}\mathcal{W}_{(q)\alpha}^{\lambda\sigma}  +\frac{2}{9}\mathcal{W}_{(s)\alpha}^{\lambda\sigma} \right) \nonumber \\
    &{}\equiv \sum_{a} \Pi^{\mu\nu\lambda\sigma}_{\textrm{OPE}(a)},
\end{align}
in agreement with the expressions derived in \cite{Colangelo:2019uex}.
The key observation by MV is that, in the chiral limit (and large-$N_c$ limit for the singlet current), the longitudinal part of \cref{eq:VVAGF} is known exactly by matching to the axial anomaly~\cite{Adler:1969gk,Bell:1969ts}. Particularly, expressing~\cite{Knecht:2003xy,Jegerlehner:2005fs}
\begin{multline}\label{eq:vva}
   \mathcal{W}_{\mu\nu\rho}(q_1,q_2) = \frac{-%\epsilon^{0123}
   1}{8\pi^2} \Big[
   -\epsilon_{\mu\nu q_1 q_2}q_{12\rho}w_L \\ +t_{\mu\nu\rho}^{(+)}w_T^{(+)}
   +t_{\mu\nu\rho}^{(-)}w_T^{(-)}
   +\tilde{t}_{\mu\nu\rho}^{(-)}\tilde{w}_T^{(-)}
   \Big],
\end{multline}
where the above form factors have an implicit dependence on momentum $w_{L,T}\equiv w_{L,T}(q_1^2,q_2^2,q_{12}^2)$, and with the tensor structures above defined as~\cite{Knecht:2003xy,Jegerlehner:2005fs}
\begin{align}
    t_{\mu\nu\rho}^{(+)} {}&= \epsilon_{q_1q_2\mu\rho}q_{1\,\nu} +\epsilon_{q_2q_1\nu\rho}q_{2\,\mu} -(q_1\cdot q_2)\epsilon_{\mu\nu\rho \bar{q}_{12}} \nonumber\\ {}& \qquad\qquad\qquad\qquad +\frac{q_1^2 +q_2^2 -q_{12}^2}{q_{12}^2}\epsilon_{\mu\nu q_1 q_2}q_{12\,\rho}, \label{eq:tP}\\
    t_{\mu\nu\rho}^{(-)} {}&= \epsilon_{\mu\nu q_1 q_2}\left[ \bar{q}_{12\,\rho} - \frac{q_1^2 -q_2^2}{q_{12}^2}q_{12\,\rho} \right], \label{eq:tM}\\
    \tilde{t}_{\mu\nu\rho}^{(-)} {}&= \epsilon_{q_1q_2\mu\rho}q_{1\,\nu} -\epsilon_{q_2q_1\nu\rho}q_{2\,\mu} -(q_1\cdot q_2)\epsilon_{\mu\nu\rho q_{12}}, \label{eq:tMT}
\end{align}
one finds via matching the axial anomaly~\cite{Adler:1969gk,Bell:1969ts}, the exact result
\begin{equation}\label{omegaL}
    w_L = 2N_c q_{12}^{-2},
\end{equation}
that must be satisfied for any value of $q_{12,3,4}^2 \ll \hat{q}^2$ and that entails nontrivial consequences for phenomenological models aiming to describe the $\langle VVA \rangle$ or the HLbL Green's functions as we shall discuss. 

While the previous decomposition in \cref{eq:vva} conveniently separates the axial anomaly, it is at the cost of introducing kinematic singularities. The fact that they are unphysical implies nontrivial relations among form factors and obscures the physical picture. Indeed, this has been a source of discrepancy among Refs.~\cite{Melnikov:2003xd,Nyffeler:2009tw,Jegerlehner:2009ry,Colangelo:2019uex,Colangelo:2019lpu} in the past. Particularly, 
for vanishing $q_{12}^2$ momentum, \cref{eq:vva} behaves as
\begin{multline} \label{eq:notunphysicalmasslesspole}
    \lim_{\substack{q_{12}^2\to0}}
    \mathcal{W}^{\mu\nu\rho}(q_1,q_2) = \lim_{q_{12}^2\to0} \frac{\epsilon^{\mu\nu q_1 q_2}q_{12}^{\rho}}{8\pi^2}\Big[  
     \frac{\operatorname{Res}(w_L)\vert_{q_{12}^2=0}}{q_{12}^2}
\\  -\frac{q_1^2 +q_2^2}{q_{12}^2}w_T^{(+)} %\\
    + \frac{q_1^2 - q_2^2}{q_{12}^2} w_T^{(-)}   \Big]
    +\mathcal{O}(q_{12}^0)\,.
\end{multline}
The $\langle VVA \rangle$ Green's function must not exhibit unphysical massless poles. These are exclusive to the lightest pseudoscalar mesons---hereinafter pseudo-Goldstone bosons (pGBs)---whose contribution $w_{L}\vert_{\textrm{pGB}}$ will be outlined in \cref{eq:pGBvva}. Cancellation of these artifact poles demands:
\begin{equation}\label{eq:nounphyspole}
 w_L = w_{L}\vert_{\textrm{pGB}} + \frac{q_1^2 +q_2^2}{q_{12}^2}w_{T0}^{(+)} - \frac{q_1^2 -q_2^2}{q_{12}^2}w_{T0}^{(-)},
\end{equation}
where we have introduced $\lim_{q_{12}^2\to0} w_T^{(\pm)}(q_1^2,q_2^2,q_{12}^2) \equiv w_{T0}^{(\pm)}(q_1^2,q_2^2)$. We note that  additional regular terms in $q_{12}^2$ in \cref{eq:nounphyspole} would contradict \cref{omegaL}.

The previous result suggests to decompose the transverse parts as
\begin{equation}\label{eq:subtway}
w_T^{(\pm)}(q_1^2,q_2^2,q_{12}^2) = w_{TS}^{(\pm)}(q_1^2,q_2^2,q_{12}^2) + w_{T0}^{(\pm)}(q_1^2,q_2^2),
\end{equation}
whose implications and possible advantages are explored in \cref{sec:axial} in the context of a phenomenological model of axial-vector mesons. We emphasize that such a decomposition is irrelevant for $\tilde{w}_T^{(-)}$. The previous equation allows to reexpress \cref{eq:vva} in the chiral limit as 
\begin{widetext}
\begin{multline}\label{eq:vvarepr}
    \mathcal{W}_{\mu\nu\rho}(q_1,q_2) = \frac{-1}{8\pi^2}\Bigg[  -\epsilon_{\mu\nu q_1 q_2}q_{12\rho}   \left( \frac{2N_c \tilde{F}_{PVV}(q_1^2,q_2^2)}{q_{12}^2 -m_{\textrm{pGB}}^2 } +\frac{(q_1^2 +q_2^2)w_{T0}^{(+)} -(q_1^2 -q_2^2)w_{T0}^{(-)}}{q_{12}^2}  \right)  \\ +t_{\mu\nu\rho}^{(+)} [ w_{TS}^{(+)} +w_{T0}^{(+)} ]
    +t_{\mu\nu\rho}^{(-)} [ w_{TS}^{(-)} +w_{T0}^{(-)} ] +\tilde{t}_{\mu\nu\rho}^{(-)} \tilde{w}_{T}^{(-)}\Bigg],
\end{multline}
\end{widetext}
where the corresponding pGB-pole contribution has been explicitly included---see \cref{sec:ppoles} for details. In particular, $\tilde{F}_{PVV}(q_1^2,q_2^2)$ is the normalized TFF ($\tilde{F}_{PVV}(0,0)=1$). The chiral limit and $m_{\textrm{pGB}}\to 0$ should be taken in \cref{eq:vvarepr}, while we explicitly include $m_{\textrm{pGB}}^2$ to emphasize its physical origin. 

\cref{eq:nounphyspole,eq:vvarepr} show what was anticipated: except for real photons, the anomaly necessitates an interplay among the pGB and transverse contributions to be satisfied \cite{Vainshtein:2002nv}. In particular, one must find
\begin{multline}\label{eq:opeCONS}
    (q_1^2 +q_2^2)w_{T0}^{(+)}(q_1^2,q_2^2)  -(q_1^2 -q_2^2)w_{T0}^{(-)}(q_1^2,q_2^2) = \\  = 2N_c[1 -\tilde{F}_{P\gamma\gamma}(q_1^2,q_2^2)].
\end{multline}
 Any model attempting to fulfill it shall specify how these TFFs are accounted for. This has profound consequences that we discuss in the following sections and make up the core of this paper.
 
 Additional interesting results are presented in the following. If we take the (symmetric) asymptotic limit in  \cref{eq:opeCONS}, 
\begin{equation}\label{eq:wTHE}
    \lim_{q^2\to \infty} w_{T0}^{(+)}(q^2,q^2)= \frac{N_c}{q^2} + \mathcal{O}(q^{-4}).
\end{equation}
It might also be interesting to consider the asymmetric case $q_2^2=0$,
\begin{equation}\label{eq:preTaylor}
    w_{T0}^{(+)}(q_1^2,0) -w_{T0}^{(-)}(q_1^2,0) = \frac{2N_c}{q_1^2}[1 -\tilde{F}_{P\gamma\gamma}(q_1^2 ,0)].
\end{equation}
In the limit $q_1^2\to 0$, and noting the TFF Taylor expansion $\tilde{F}_{P\gamma\gamma} = 1 +b_{\pi}q_1^2 + ...$, one finds the corresponding leading term for $w_{T0}^{(+)}(q_1^2,0)$  
\begin{equation}\label{eq:pi0slope}
w_{T0}^{(+)}(q_1^2,0) = -2N_c b_{\pi} + \mathcal{O}(q_1^2).   
\end{equation}
 Remarkably, this naturally explains the nonperturbative corrections to the nonrenormalization theorems found in \cite{Knecht:2003xy} (see Eqs.~(3.5) and (3.6) therein and Ref.~\cite{Vainshtein:2002nv}) without the need to resort to chiral perturbation theory. Still in the asymmetric case, but in the high-energy extreme, for asymptotically large $q_1^2$, using \cref{eq:ptffNORM} and the Brodsky-Lepage limit \cite{Lepage:1979zb, Lepage:1980fj} [except for a factor 3 this is essentially  \cref{eq:ptffOPE} below]
\begin{multline} \label{eq:omegasdiff}
    w_{T0}^{(+)}(q_1^2,0) -w_{T0}^{(-)}(q_1^2,0)  \\ = \frac{2N_c}{q_1^2}\left[ 1 +\frac{24\pi^2 F_P^2}{N_c q_1^2} \right] + \mathcal{O}(q_1^{-6}).
\end{multline}
We stress that previous results,  \cref{omegaL,eq:notunphysicalmasslesspole,eq:nounphyspole,eq:subtway,eq:vvarepr,eq:opeCONS,eq:wTHE,eq:preTaylor,eq:pi0slope,eq:omegasdiff}, hold only in the chiral (and large-$N_c$ for $a=0$) limit. 

Concluding this section, we have shown the interplay between the transverse and pGB-pole contributions, and its particular manifestation for symmetric and antisymmetric FFs. In the next section we shall explore the role of pGBs in \cref{eq:opeCONS}.

\section{\label{sec:ppoles}Pseudoscalar-poles contribution}

In this section, we provide the relevant results concerning the lightest pseudoscalar-poles (e.g. pGBs) contributions to the HLbL and $\langle VVA \rangle$ Green's functions, including the difficulties to fulfill $\langle VVA \rangle$ constraints with pseudoscalar contributions. To that end, we introduce the pseudoscalar TFFs
\begin{multline}\label{eq:PTFF}
    i\int d^4x e^{i q_1\cdot x} \bra{0} T\{ j^{\mu}(x) j^{\nu}(0) \} \ket{P(q_{12})} \\ = \epsilon^{\mu\nu q_1 q_2}F_{P\gamma\gamma}(q_1^2,q_2^2),
\end{multline}
whose high-energy behavior can also be derived using the OPE,  \cref{eq:opegen},
\begin{equation}\label{eq:ptffOPE}
    \lim_{q^2\to\infty}F_{P\gamma\gamma}(q^2,q^2) = -\frac{1}{\hat{q}^2}\sum_c 2F_P^{c} \operatorname{tr}(\lambda^c \mathcal{Q}^2),
\end{equation}
where $\bra{0} A_{\mu}^c \ket{P(q)} = iq_{\mu}F_P^c $.  Further, in the chiral (and large-$N_c$ for $c=0$) limit, the normalization for real photons is fixed as well. Particularly~\cite{Adler:1969gk, Bell:1969ts, Bardeen:1969md},
\begin{equation}\label{eq:ptffNORM}
    \sum_P F_P^c F_{P\gamma\gamma}(0,0) = \frac{N_c}{4\pi^2}\operatorname{tr}(\lambda^c \mathcal{Q}^2).
\end{equation}
Note that, if the electromagnetic current is replaced as $j_{\mu(\nu)} \to V_{\mu(\nu)}^{a(b)}$, the TFF in \cref{eq:PTFF} is relabelled following $F_{P\gamma\gamma}(q_1^2,q_2^2) \to F_{PV_aV_b}(q_1^2,q_2^2)$, while \cref{eq:ptffOPE,eq:ptffNORM} would be modified as $\operatorname{tr}(\lambda^c \mathcal{Q}^2) \to \operatorname{tr}(t^c \{ t^a,t^b \})$. With these definitions, it is straightforward to obtain the pGBs-pole contribution to the $\langle VVA \rangle$ and HLbL Green's functions. For the $\langle VVA \rangle$, one obtains the single pGB-pole contribution to \cref{eq:VVAGF}
\begin{equation}\label{eq24}
\mathcal{W}^{\mu\nu\rho}_{abc}(q_1,q_2)|_P = \epsilon^{\mu\nu q_1 q_2}q_{12}^{\rho} \frac{F_P^cF_{PV_aV_b}(q_1^2,q_2^2)}{q_{12}^2 -m_{\textrm{pGB}}^2}. 
\end{equation}
Specializing to the chiral limit, and making use of \cref{eq:ptffNORM} to isolate $F_P^c$, it is straightforward to extract the pGB-poles contribution to $w_L$, that we named $w_L|_{\textrm{pGB}}$ in \cref{eq:nounphyspole} after comparing \cref{eq24} with \cref{eq:notunphysicalmasslesspole}:
\begin{equation}\label{eq:pGBvva}
 w_L|_{\textrm{pGB}} = \frac{2N_c \tilde{F}_{PVV}(q_1^2,q_2^2)}{q_{12}^2 -m_{\textrm{pGB}}^2},
\end{equation}
where the limit $m_{\textrm{pGB}}\to 0$ shall be taken, though it is kept in this equation to emphasize its physical origin. In the equation above, we have introduced the normalized TFF $\tilde{F}_{PVV}(q_1^2,q_2^2)\equiv F_{PVV}(q_1^2,q_2^2)/F_{PVV}(0,0)$. Note that, in the chiral limit (and large-$N_c$ limit for the singlet current), $U(3)$ symmetry implies the same $q_{1,2}^2$-dependence for any current.

Regarding the HLbL, the individual pGB-pole contributions reads
\begin{multline}
  \Pi^{\mu\nu\lambda\sigma}(q_1,q_2,q_3) = 
  \frac{-i \epsilon^{\mu\nu q_1 q_2} \epsilon^{\lambda\sigma q_3 (-q_4)} }{q_{12}^2 -m_{P}^2}  \\ \times F_{P\gamma\gamma}(q_1^2,q_2^2)F_{P\gamma\gamma}(q_3^2,q_4^2)  + (t, u),
\end{multline}
where the last term stands for the $t$ and $u$ channel contributions. Let us now scrutinize its behavior in the energy regime relevant to the OPE constraint, \cref{eq:hlblope}. Defining $\tilde{F}_{P\gamma\gamma}(q_3^2,q_4^2) \equiv F_{P\gamma\gamma}(q_3^2,q_4^2)/F_{P\gamma\gamma}(0,0)$, together with \cref{eq:ptffOPE} and \cref{eq:ptffNORM}, and particularizing to the $\pi^0$ case, its pole behaves as
\begin{equation}\label{eq:hlblpi0}
    \Pi^{\mu\nu\lambda\sigma}_{\textrm{OPE}} (q_1,q_2,q_3) |_{\pi^0}=
    \frac{i \epsilon^{\mu\nu q_1 q_2} \epsilon^{\lambda\sigma q_3 (-q_4)} }{6\pi^2\hat{q}^2(q_{12}^2 -m_{\pi}^2)} \tilde{F}_{\pi\gamma\gamma}(q_3^2,q_4^2),
\end{equation}
where the chiral limit is finally obtained by taking $m_{\pi}^2\to0$. This shall be compared to the OPE constraint for the isotriplet $w_L$ component in \cref{eq:hlblope,eq:opeisospin}. Using \cref{omegaL}, such constraint reads
\begin{equation}\label{eq:ope3}
\lim_{w_T^{(\pm)},\tilde{w}_T^{(-)}\to0}\Pi^{\mu\nu\lambda\sigma}_{\textrm{OPE}(3)} (q_1,q_2,q_3)  = 
    \frac{i \epsilon^{\mu\nu q_1 q_2} \epsilon^{\lambda\sigma q_3 (-q_4)} }{6\pi^2 \hat{q}^2 q_{12}^2}\,.
\end{equation}
Consequently, \cref{eq:hlblpi0} satisfies the anomaly in the chiral limit for real photons while, for virtual photons, the $q_{3,4}^2$-dependence induced by $\tilde{F}_{\pi\gamma\gamma}(q_3^2,q_4^2)$ demands the interplay with transverse contributions in order to comply with the anomaly,  as shown in \cref{eq:opeCONS}, which cannot be reproduced with modified form factors or extended pseudoscalar sectors \cite{Nyffeler:2009tw, Colangelo:2019uex, Colangelo:2019lpu}.

\section{\label{sec:axial}Implications for axial-vector mesons}

\subsection{General discussion}

In the following, we present a phenomenological model for axial-vector-mesons contributions that will help illustrating the comments above, and shows how axial-vector mesons might conspire, together with the pseudoscalar poles in the previous section, to fulfill \cref{eq:opeCONS}. First, we introduce the general definitions and discuss these contributions to the $\langle VVA \rangle$ Green's function for simplicity, continuing to the HLbL case in \cref{sec:axialsHLBL}. The first ingredient we need is the axial-vector meson TFF,
\begin{multline}
   i \int d^4 x e^{iq_1\cdot x} \bra{0} T\{ j^{\mu}(x) j^{\nu}(0) \}  \ket{A(q_{12})} \\ = \mathcal{M}_A^{\mu\nu\rho}(q_1,q_2)\varepsilon_{A\rho}. \label{eq:ATFF}
\end{multline}
To describe $\mathcal{M}_A^{\mu\nu\rho}(q_1,q_2)$, we choose the basis in Ref.~\cite{Roig:2019reh}, that we will find to be optimal for reconstructing the axial-vector-mesons contributions to the $\langle VVA \rangle$ and HLbL Green's functions. This reads
\begin{multline}\label{eq:axialFF}
    \mathcal{M}_A^{\mu\nu\rho}(q_1,q_2)=
    i\Big[ \epsilon^{\mu\nu q_1 q_2}q_{12}^{\rho}C_S +\epsilon^{\mu\alpha\rho q_1}(q_{2\,\alpha}q_2^{\nu} -g^\nu_\alpha q_2^2) B_2\\
    +\epsilon^{\nu\alpha\rho q_2}(q_{1\,\alpha}q_1^{\mu} -g^\mu_\alpha q_1^2) \bar{B}_2  
    +\epsilon^{\mu\nu q_1q_2} \bar{q}_{12}^{\rho}C_A \Big],
\end{multline}
with $\bar{B}_2(q_1^2,q_2^2) = B_2(q_2^2,q_1^2)$ and $C_{A}(q_1^2,q_2^2)$ symmetric under the exchange $(q_1\leftrightarrow q_2)$. The $C_S(q_1^2,q_2^2)$ form factor is symmetric and unphysical ($q_{12}\cdot\varepsilon_A(q_{12})=0$), but necessary to keep results basis independent~\cite{Roig:2019reh} and to provide a basis for the $\langle VVA \rangle$ Green's function free of kinematic singularities.
Also, we will need the doubly-virtual high-energy behavior, that can be obtained by means of the OPE. Inserting the result from \cref{eq:opegen} into \cref{eq:ATFF}, and taking the corresponding limit in \cref{eq:axialFF}, one obtains~\cite{Roig:2019reh}\footnote{
We take $\bra{0}A_{\mu}^a \ket{A(q)} \equiv F_A m_A \varepsilon_{A\mu}$, with nonvanishing $F_{f_1}^q= F_{f_1'}^s= F_{a_1}^3= F_A \in [130,150]$~MeV~\cite{Dumm:2009va, Nugent:2013hxa}, while the Weinberg sum rules~\cite{Weinberg:1967kj} imply $F_A\simeq \sqrt{2}F_{\pi}$~\cite{Roig:2013baa}.}
\begin{equation}\label{eq:OPEaxialTFF}
    \lim_{\hat{q}^2\to\infty}B_{2S}(\hat{q}^2,\hat{q}^2)\varepsilon_A^{\rho} = \frac{1}{\hat{q}^4}\bra{0} j_5^{\rho} \ket{A} 
   \equiv \sum_a \frac{m_A F_A^a}{\hat{q}^4}\varepsilon_A^{\rho}\operatorname{tr}\mathcal{Q}^2\lambda^a,
\end{equation}
where we have introduced $B_{2S} = (B_2 +\bar{B}_2)/2$, see also Ref.~\cite{Hoferichter:2020lap}.  
To arrive to this expression, we used the power counting related to $q_1^2\sim q_2^2\sim \hat{q}^2 \gg q_{12}^2$. 

Indeed, it is useful to express $B_2$ and $\bar{B}_2$ in terms of form factors with well-defined symmetry under $(q_1\leftrightarrow q_2)$, $B_2 = B_{2S} +B_{2A}$ and $\bar{B}_2 = B_{2S} -B_{2A}$, that allows to 
rewrite \cref{eq:axialFF}  in terms of the tensor structures in \cref{eq:tP,eq:tM,eq:tMT} as 
\begin{multline}\label{eq:axialTFFvvabasis}
    \mathcal{M}^A_{\mu\nu\rho} = i\Bigg\{  t_{\mu\nu\rho}^{(+)}(-B_{2S}) + t_{\mu\nu\rho}^{(-)}(C_A -B_{2A}) + \tilde{t}_{\mu\nu\rho}^{(-)}(B_{2A}) \\
    -\epsilon_{\mu\nu q_1 q_2}q_{12\rho} \Big[-C_S -\frac{q_1^2+q_2^2}{q_{12}^2}B_{2S} -\frac{q_1^2-q_2^2}{q_{12}^2}(C_{A} -B_{2A}) \Big]\Bigg\}.
\end{multline}
From  \cref{eq:axialTFFvvabasis} it is clear that, whenever $q_{12}^{\rho}\mathcal{M}^A_{\mu\nu\rho} \neq 0$ [that corresponds precisely to the last line in \cref{eq:axialTFFvvabasis}], the axial-vector-mesons contributions to the $\langle VVA \rangle$ or the HLbL Green's functions will feature apparent unphysical massless-pole contributions to $w_L$. %Such contributions
These are key to satisfy the anomaly constraint in the form of \cref{eq:opeCONS}. 

To illustrate this, let us focus first on the $\langle VVA \rangle$ Green's function. Decomposing into isospin channels as in \cref{eq:opeisospin}, and specializing to electromagnetic currents, the axial-vector-mesons contributions to the $\langle VVA \rangle$ Green's function read
\begin{align}\label{eq:AxialsVVA}
    \mathcal{W}_{\mu\nu\rho}^a\vert_A {}&= \sum_{A,\textrm{pol.}}i\int d^4x e^{i q_1\cdot x} \bra{0} T\{ j_{\mu}(x) j_{\nu}(0) \} \ket{A(q_{12})} \nonumber\\ 
    & \qquad\qquad\qquad
    \times \frac{i}{q_{12}^2 -m_A^2} \bra{A(q_{12})} A^a_{\rho} \ket{0} \nonumber \\ {}&=
    i \sum_{A,\textrm{pol.}} \mathcal{M}_{\mu\nu\tau}^A m_A F_A^a 
    \frac{\varepsilon_A^{\tau}\varepsilon_{A\rho}^*}{q_{12}^2 -m_A^2}.
\end{align}
Noting the axial propagator
\begin{equation}\label{eq:axialpropSTD}
    D^{\tau\rho}(q_{12}^2) \equiv \frac{\sum_{\textrm{pol}} \varepsilon_A^{\tau}\varepsilon_A^{\rho*}}{q_{12}^2 -m_A^2} = -\frac{g^{\tau\rho} -\frac{q_{12}^{\tau}q_{12}^{\rho}}{m_A^2}}{q_{12}^2 -m_A^2},
\end{equation}
and using the results in \cref{eq:axialTFFvvabasis}, we find easily the individual axial-vector-mesons contributions in \cref{eq:AxialsVVA}. Expressed in the basis of \cref{eq:vva}, the relevant form factors read
\begin{multline}
  \frac{ \{ w_T^{(+)} , w_T^{(-)} , \tilde{w}_T^{(-)}  \} }{8\pi^2} = \frac{ \{ B_{2S}, B_{2A} -C_{A}, -B_{2A}  \}}{q_{12}^2 -m_A^2}m_A F_A^a, \\
  \frac{w_L}{ 8\pi^2 } = \Big[C_S +\frac{q_1^2+q_2^2}{q_{12}^2}B_{2S} -\frac{q_1^2-q_2^2}{q_{12}^2}(B_{2A} -C_{A}) \Big]
  \\ \times \left[1 -\frac{q_{12}^2}{m_A^2}\right]\frac{m_A F_A^a}{q_{12}^2 -m_A^2}, \label{eq:vvaALSTD}
\end{multline}
confirming that, in general, axial-vector mesons induce a nonvanishing $w_L$ contribution.  This is precisely the kind of contribution required in \cref{eq:opeCONS} to fulfill the anomaly. In particular, adding the pGB contribution to the $\langle VVA \rangle$ Green's function (see \cref{sec:ppoles}), one finds
\begin{multline}
  \frac{w_L N_c \operatorname{tr}(\mathcal{Q}^2\lambda^a)}{8\pi^2} = \frac{N_c \operatorname{tr}(\mathcal{Q}^2\lambda^a)}{4\pi^2}\frac{\tilde{F}_{P\gamma\gamma}(q_1^2,q_2^2)}{q_{12}^2 -m_{\textrm{pGB}}^2}  -\sum_n \frac{F_{A_n}^a}{m_{A_n}^a}\Big[C_S  \\ +\frac{q_1^2+q_2^2}{q_{12}^2}B_{2S} -\frac{q_1^2-q_2^2}{q_{12}^2}(B_{2A} -C_{A}) \Big] = \frac{N_c \operatorname{tr}(\mathcal{Q}^2\lambda^a)}{4\pi^2 q_{12}^2}, \label{eq:modelVVAano}
\end{multline}
that is the analogous of \cref{eq:opeCONS}. When accounting for the trace factors, the anomaly condition \cref{omegaL} yields the last equality and, together with \cref{eq:opeCONS}, it requires a vanishing $C_S=0$ form factor in order to accommodate the appropriate $q_{12}^2$ dependence. This justifies to drop such an unphysical (purely longitudinal) form factor, and makes this basis \cref{eq:vva} ideal. While this might seem an obvious choice, this feature is peculiar to this basis and connects to the problem of the basis ambiguities first outlined in Ref.~\cite{Roig:2019reh} if longitudinal form factors are omitted. To illustrate this, we take the basis in Ref.~\cite{Knecht:2020xyr}, whose form factors ($w_{0,1,2,3}$) are related to ours ($C_S,C_A,B_{2S},B_{2A}$) via Schouten identities:
\begin{align}\label{eq:basistranslate}
    w_0 &= C_S +B_{2S} & w_1 &= C_A & w_2 &= -B_{2S} & w_3 &= B_{2A}\, , 
\end{align}
\noindent
which can also be reverted to
\begin{align}\label{eq:basistranslate2}
    C_S &= w_0 +w_2 & C_A &= w_1 & B_{2S}&= -w_2 & B_{2A} &= w_3.
\end{align}
If its unphysical (purely longitudinal) form factor $w_0$ is discarded (e.g. $w_0=0$ is taken), a nonvanishing $B_{2S}=-w_2$ form factor necessarily implies a nonvanishing longitudinal $C_S=w_2=-B_{2S}$ form factor, in violence with the anomaly as previously discussed. Hence, dropping longitudinal form factors is not necessarily justified in every basis. This has dramatic consequences as we will illustrate in the following section in the context of the HLbL. How could one fix the unphysical part? As we found, the anomaly offers an answer for 
every basis. In the previous case, it would require $w_0=-w_2$ instead of $w_0 = 0$; in our basis, it simply demands $C_S=0$. This method to fix the unphysical form factor (and to overcome the basis ambiguities outlined in Ref.~\cite{Roig:2019reh}) is, to our best knowledge, a novel result. In practice, we recommend to use the basis in \cref{eq:axialFF} with $C_S=0$ for simplicity. Otherwise, unphysical longitudinal form factors will be required.  
\\

An important final remark is that axial-vector-mesons contributions to $w_L$ lack an axial pole, meaning they are unphysical (they cannot be unambiguously attributed to axial-vector mesons). As an example, in resonance chiral theory (RChT) \cite{Ecker:1988te}, axial-vector mesons produce purely transverse contributions---terms akin to $w_L$ in \cref{eq:vvaALSTD} arise from contact terms (see for instance Ref.~\cite{Kadavy:2016lys}). To better understand this, we split the propagator as follows
\begin{multline}\label{eq:axialPROP}
D^{\alpha\beta}(q^2) =
    -\frac{q^2g^{\alpha\beta} -q^{\alpha}q^{\beta}}{m_A^2(q^2 -m_A^2)} +\frac{g^{\alpha\beta}}{m_A^2} \equiv \bar{D}^{\alpha\beta}(q^2) +\frac{g^{\alpha\beta}}{m_A^2}.
\end{multline}
Note that $\bar{D}^{\alpha\beta}(q^2)$ induces purely transverse terms, vanishing at $q^2=0$ and with an axial pole; these contribute to what we dubbed as $w_{TS}^{(\pm)}$ in \cref{eq:vvarepr}. Conversely, the  $g^{\alpha\beta}m_A^{-2}$ term has no pole or any $q^2$ dependence at all; it contributes to what we named $w_{T0}^{(\pm)}$ in \cref{eq:vvarepr}. Recall that, together with the pGB-pole contribution, such term is responsible for fulfilling the anomaly, cf. \cref{eq:opeCONS}. The whole resummation of axial-vector mesons could then be expressed in terms of the pGB TFF. This offers a suggestive possibility: to consider physical axial-vector-mesons contributions using $\bar{D}^{\alpha\beta}(q^2)$ in a phenomenological manner, while accounting for the remaining part using a model that (together with the pGBs) satisfies the anomaly. While this is simple for the $\langle VVA \rangle$ case, the HLbL demands satisfying both, \cref{eq:hlblope}, and \cref{eq:opeCONS}. The required asymptotics forbid a simple model and suggest the necessity of an infinite number of terms. 

In the following section, we discuss the role of axial-vector mesons in the context of the HLbL contribution to the anomalous magnetic moment of the muon, $a_{\mu}^{\textrm{HLbL;}A}$. In particular, we review the current status, addressing the implications of the high-energy behavior \cref{eq:opegen}, the relevance of choosing an appropriate basis, and comment on the possible splitting of their contributions according to \cref{eq:axialPROP}.

\subsection{The role of axial-vector mesons in the HLbL\label{sec:axialsHLBL}}

The axial-vector-mesons contribution to the HLbL tensor can be expressed generically as 
\begin{multline}
    \Pi_{\mu\nu\lambda\sigma}(q_1,q_2,q_3) = -i\mathcal{M}^A_{\mu\nu\alpha}(q_1,q_2) D^{\alpha\beta}(q_{12}) \\ \times \mathcal{M}^{A}_{\lambda\sigma\beta}(q_3,-q_4) +(t,u).
\end{multline}
where the last term stands for the $t$ and $u$ channel contributions. In the following, we focus on $B_{2S}$ form factor, since it provides the bulk contribution and it is the only one surviving in the kinematic limit studied with the help of the OPE in \cref{eq:OPE1,eq:hlblope}, see \cref{eq:OPEaxialTFF}. Besides, this is the only form factor for which experimental data exists. In particular, this is available for the $f_1$ and $f_1'$ mesons for a singly-virtual photon~\cite{Achard:2001uu,Achard:2007hm}. In the past, the dipole parametrization used in \cite{Achard:2001uu,Achard:2007hm} has been extended to the doubly-virtual regime in Ref.~\cite{Pauk:2014rta} as 
\begin{equation}
B_{2S,A}^{\textrm{Fact}}(q_1^2,q_2^2) = \frac{B_{2S}^A(0,0)\Lambda_A^8}{(\Lambda_A^2 -q_1^2)^2(\Lambda_A^2 -q_2^2)^2},
\end{equation}
that we will refer to as the \textit{Fact} ansatz (stemming from \textit{factorized} ansatz). Note that the result in Ref.~\cite{Pauk:2014rta} using this model is amongst the currently accepted values for $a_{\mu}^{\textrm{HLbL;}A}$ in Ref.~\cite{Aoyama:2020ynm}. However, if compared to \cref{eq:OPEaxialTFF}, it is clear that its doubly-virtual high-energy behavior is incorrect, and so will be the scaling for \cref{eq:OPE1}. To amend this, hereinafter we adopt the following ansatz \cite{Masjuan:2015lca}
\begin{equation}\label{eq:axialB2Sope}
B_{2S,A}^{\textrm{OPE}}(q_1^2,q_2^2) = \frac{B_{2S}^A(0,0)\Lambda_A^4}{(\Lambda_A^2 -q_1^2 -q_2^2)^2},
\end{equation}
that has the same singly-virtual behavior but the correct high-energy scaling. We will refer to it as the \textit{OPE} ansatz. Before continuing our discussion, it is relevant to emphasize that the induced difference is by no means small. To illustrate this, we show in \cref{tab:amuAx} the individual axial-vector-mesons contributions to $a_{\mu}^{\textrm{HLbL;}A}$ using either the \textit{Fact} or \textit{OPE} ansatz.\footnote{Noting that $B_{2S}^{A}(0,0) = [12 \tilde{\Gamma}_{\gamma\gamma}^A/ (\pi \alpha^2 m_A^5)]^{1/2}$~\cite{Roig:2019reh},  Refs.~\cite{Achard:2001uu,Achard:2007hm} imply $B_{2S}^{f_1,f_1'}(0,0)=\{0.269(30),0.197(30)\}~\textrm{GeV}^{-2}$, while $\Lambda_{f_1,f_1'} = \{1.04(8),0.926(79)\}$~GeV. For the $a_1$, we take $B_{2S}^{a_1}(0,0)=0.245(63)~\textrm{GeV}^{-2}$ and $\Lambda_{a_1} = 1.0(1)$~GeV~\cite{Roig:2019reh}.} Implementing an appropriate high-energy behavior roughly doubles the value obtained with the \textit{Fact} ansatz, that points to a current underestimation of $a_{\mu}^{\textrm{HLbL;}A}$. Note also that this result is in line with the holographic ones in Ref.~\cite{Leutgeb:2019gbz} for the first axial-vector meson multiplet, that also have an appropriate description for the high-energy behavior and yield nonvanishing $B_{2S},B_{2A}$ TFFs.

\begin{table}[tbp]
\begin{tabular}{ccccc} \hline
   & $f_1$ & $f_1'$ & $a_1$ & Total \\ \hline
Fact & $4.3(^{+1.8}_{-1.5})$ &  $1.2(^{+0.6}_{-0.5})$ & $2.8(^{+1.9}_{-1.7})$ & $8.3(^{+2.7}_{-3.4})$ \\
OPE & $8.3(^{+3.4}_{-2.9})$ &  $2.3(^{+1.1}_{-0.9})$ & $5.4(^{+3.7}_{-3.3})$ & $16.0(^{+5.1}_{-4.5})$ \\ \hline
\end{tabular}
\caption{\label{tab:amuAx}Individual axial-vector-mesons contributions to $a_{\mu}^{\textrm{HLbL;}A}$ from $B_{2S}$ form factor only. The second row corresponds to the {\textit{factorized}} FF and is intended for illustration. The third row corresponds to the {\textit{OPE}} FF, that has the correct high-energy scaling.}
\end{table}

Now, we come back to the important point of fixing  potential unphysical longitudinal form factors. In the case above, we disregarded the $C_S=0$ form factor that, as we argued in the previous section, was justified on the basis of the anomaly. If, instead, we would use again the basis from Ref.~\cite{Knecht:2020xyr} [see \cref{eq:basistranslate}], the experimental data would constraint now $w_2$ while, naively, one would drop the longitudinal form factor $w_0$. Proceeding this way, and choosing $w_{2} = B_{2S}^{\textrm{OPE}}$, one would obtain $a_{\mu}^{f_1,f_1',a_1} = \{3.48, 1.01, 2.25\}\times 10^{-11}$, half the value in our basis, see last row in \cref{tab:amuAx}. This illustrates the effects of the basis dependence (if longitudinal form factors are dropped) in the context of $a_{\mu}$ that, to our best knowledge, was unnoticed before. To understand this, we shall recall what we discussed after \cref{eq:basistranslate}: choosing $w_{2} = -B_{2S}^{\textrm{OPE}}$ and $w_0=0$ implies a nonvanishing $C_S = -B_{2S}^{\textrm{OPE}}$ in our basis. To comply with the anomaly, we showed that $w_0=-w_2$ should have been taken instead, rendering $C_S=0$ and recovering our results. As we said, in order to avoid the need for including nonvanishing unphysical form factors, we recommend to stick to the basis in \cref{eq:axialFF}, which is one of the important messages of this work. Current results restricting to a nonvanishing $B_{2S}$ TFF~\cite{Pauk:2014rta} (that is due to the limited available phenomenological information) are unaffected by this issue; while most bases proposed to tackle all the TFFs~\cite{Jegerlehner:2015stw,Pascalutsa:2012pr} are.

Finally, we come back to the possibility to split the axial-vector-mesons contributions to $a_{\mu}$ according to \cref{eq:axialPROP}. As stated, the part related to $\bar{D}^{\alpha\beta}(q^2)$ will produce purely transverse contributions vanishing at $q^2=0$ and such vanishing renders them numerically subleading as well. For obvious reasons, we will call them the \textit{subtracted} part; the part related to $g^{\alpha\beta}$ in turn will be $q^2$ independent, and we will refer to it as a \textit{contact} part.  In \cref{tab:amuAxCS}, we show the individual axial-vector-mesons contributions split into such subtracted and contact parts that, of course, add up to the results in \cref{tab:amuAx}, last row. Clearly, the contact part provides the bulk of the contribution, while the subtracted one is smaller and with opposite sign. Noteworthy, this explains the results obtained in RChT~\cite{Roig:2019reh} for the symmetric $B_{2S}$ form factor---smaller and with opposite sign as compared to the estimates in \cite{Pauk:2014rta,Jegerlehner:2017gek}. The reason is simple: RChT provides purely transverse results akin to the subtracted contributions for $B_{2S}$. As an interesting outcome of our results, this suggests that, if one could create a reliable model for the contact part for the full tower of axial-vector mesons required to fulfill the anomaly, most of the current error ascribed to the TFFs would be greatly reduced. We further speculate on this in \cref{sec:simplemodel}.

\begin{table}[tbp]
\begin{tabular}{ccccc} \hline
   & $f_1$ & $f_1'$ & $a_1$ & Total \\ \hline
Subt & $-1.3(^{+0.5}_{-0.7})$ &  $-0.27(^{+0.12}_{-0.17})$ & $-0.85(^{+0.55}_{-0.68})$ & $-2.42(^{+0.75}_{-1.0})$ \\
Cont & $9.6(^{+4.1}_{-3.5})$ &  $2.6(^{+1.3}_{-1.0})$ & $6.2(^{+4.3}_{-3.8})$ & $18.4(^{+6.1}_{-5.3})$ \\ \hline
\end{tabular}
\caption{\label{tab:amuAxCS} The results of {\textit{OPE}} row in \cref{tab:amuAx} if split into subtracted and contact parts. Their sum matches the result in \cref{tab:amuAx}, {\textit{OPE}} row.}
\end{table}

In summary, the currently accepted value for $a_{\mu}^{\textrm{HLbL;}A}$ stemming from the first axial-vector meson multiplet, ($B_{2S}$ form factor contribution only)  should be increased due to the OPE behavior up to $16.0(^{+5.1}_{-4.5})\times 10^{-11}$. The difference may then be attributed to a systematic error which is lacking so far. Beyond that, experimental data is also required in order to pin down the errors, especially regarding the so far missing $a_1$ resonance. More importantly, considering the full axial-vector-mesons contributions requires setting a basis. However, as first outlined in Ref.~\cite{Roig:2019reh} and here further developed, ambiguities arise in connection to the longitudinal form factor and the usage of Schouten identities. We have illustrated its impact and found a solution to this problem. In particular, our basis (with $C_S=0$) should be used, or unphysical form factors be incorporated if necessary. Last, we have discussed the possibility to split the contribution into a subtracted and a contact part. This opens the possibility to use a model convoluted with the contact part complying with the anomaly (see \cref{sec:simplemodel}) and also explains the results found in Ref.~\cite{Roig:2019reh} using RChT. While these are our main results, to complete our study, we discuss in the following subsection a further interesting application. Last, in \cref{sec:simplemodel}, we illustrate our comments with the aid of a simple model with infinite resonances as well as using the holographic model from Ref.~\cite{Leutgeb:2019gbz}.

\subsection{\label{sec:missing}Estimating  missing degrees of freedom}

As a further application, we discuss in this subsection the possibility of using \cref{eq:opeCONS} as a sum rule in the phenomenological model of pGB and axial-vector mesons  introduced in \cref{sec:axial}. Restricting to the isotriplet contributions and using \cref{eq:ptffNORM}, one finds that \cref{eq:opeCONS} [or, equivalently, \cref{eq:modelVVAano}], reads in the chiral limit
\begin{multline}\label{eq:sumrulemodel}
  F_{\pi}F_{\pi\gamma\gamma}(q_1^2,q_2^2) -\frac{N_c\operatorname{tr}(\lambda^3 \mathcal{Q}^2)}{4\pi^2} =  \sum_{n} 
\Big[ (q_1^2 +q_2^2)B_{2S}^{a_1}(q_1^2,q_2^2) \\ + (q_1^2 -q_2^2)\{ C_A^{a_1}(q_1^2,q_2^2) -B_{2A}^{a_1}(q_1^2,q_2^2) \} \Big]\frac{F_{a_1}}{m^{a_1}_n},
\end{multline}
where summation is over the whole tower of isotriplet axial-vector mesons. Such a sum rule can be used to constrain the parameters of as yet unmeasured resonance properties. For instance, expanding the left hand side in powers of $q_{i}^2$, the term of order $\mathcal{O}(q_i^2)$ implies [see comments below \cref{eq:preTaylor}]
\begin{equation}\label{eq:SRO0}
\frac{b_{\pi}N_c \operatorname{tr}(\lambda^3 \mathcal{Q}^2)}{4\pi^2} = \sum_{a_1}\frac{F_{a_1}}{m_{a_1}}B_{2S}^{a_1}(0,0).
\end{equation}
Employing $b_{\pi}= 1.76(11)$~GeV$^{-2}$ from Ref.~\cite{Masjuan:2017tvw}, obtained from PAs (see also Ref.~\cite{Hoferichter:2018kwz}), $N_c\operatorname{tr}(\lambda^3 \mathcal{Q}^2) = 1$ and $B_{2S}^{a_1}(0,0) = 0.245(63)$GeV$^{-2}$ (see footnote~2), we find $0.045(3)$~GeV$^{-2}$ for the left-hand side of \cref{eq:SRO0}, while the $a_1(1260)$ contribution yields $0.027(7)$~GeV$^{-2}$ [using the half-width rule \cite{Masjuan:2012gc,Masjuan:2012sk} to estimate further uncertainties yields an additional $(^{+0.009}_{-0.005})$ error]. We used $m_{a_1}=1209(4)$~MeV~\cite{Mikhasenko:2018bzm} and $F_{a_1}=140(10)$~MeV~\cite{Dumm:2009va, Nugent:2013hxa}. This shows that the $a_1(1260)$ contribution dominates over additional resonances at low energies. In order to estimate heavier axial-vector-mesons effects, one might assume the whole summation in \cref{eq:SRO0} to be saturated by $a_1(1260)$ and $a_1(1640)$ at low energies. This implies $\frac{F_{a_1(1640)}}{m_{a_1(1640)}}B_{2S}^{a_1(1640)}(0,0) =0.017(8)$~GeV$^{-2}$. Assuming that $F_{a_1(1260)} = F_{a_1(1640)}$~\cite{Afonin:2004yb}, we obtain $B_{2S}^{a_1(1640)}(0,0) =0.20(10)$~GeV$^{-2}$, that could be considered as a sort of upper bound. 

While the results above have been obtained in the chiral limit, $m_{\pi}^2$ corrections are expected to be small. To start with, those affecting the pion propagator can be rearranged as 
\begin{equation}
\frac{1}{q^2} \to \frac{1}{q^2} + \frac{m_{\pi}^2}{q^2(q^2 -m_{\pi}^2)}.
\end{equation}
The first part corresponds to the chiral limit and leads, once again, to \cref{eq:sumrulemodel} upon matching to the anomaly; the second part is a remainder, that is related to the $\langle VVP \rangle$ (vector-vector-pseudoscalar three-point) Green's function as a result of $\partial^{\mu}A_{\mu}^3 = \hat{m}\bar{q}\lambda^3 i\gamma^5 q$, and does not affect our discussion in this section (see comments later). Continuing with chiral corrections beyond the pion pole, $m_{\pi}^2$ corrections to  $F_{\pi\gamma\gamma}(0,0)$ are small, as measured experimentally in PrimEx-II~\cite{PrimEx2}. Similarly, $m_{\pi}^2$ corrections to $b_{\pi}$ should be small: since the slope is dominated by the $\rho,\omega$ peaks, its shift would be related to that of $m_{\rho,\omega}$ away from the chiral limit, that is negligible (see Appendix D from Ref.~\cite{Roig:2019reh} and references therein). 

Quite similarly, previous results could be used to constrain the slope of the axial-vector mesons TFFs. Regarding $B_{2S}$, that provides with the dominant contribution to $a_{\mu}^{\textrm{HLbL}}$, taking the symmetric kinematics $q_1^2 = q_2^2$, and defining
\begin{equation}
\tilde{F}_{\pi\gamma\gamma}(q_1^2,q_2^2) = 1 +b_{\pi}(q_1^2 +q_2^2) +c_{\pi}(q_1^4 +q_2^4) + a_{\pi;1,1}q_1^2 q_2^2 + ... 
\end{equation}
yields, for our TFF parametrization,
\begin{equation}\label{eq:SRO1}
\frac{2c_{\pi} +a_{\pi;1,1}}{4\pi^2} = \sum_{a_1}\frac{F_{a_1}}{m_{a_1}}B_{2S}^{a_1}(0,0)\frac{4}{\Lambda^2_{a_1}}.
\end{equation}
Using the results in Ref.~\cite{Masjuan:2017tvw}, $c_{\pi}=3.13(66)$~GeV$^{-4}$ and $a_{\pi;1,1}= \in (1.9,2.1)b_{\pi}^2$, we obtain for the left-hand side $0.32(5)$~GeV$^{-4}$, while for the $a_1(1260)$ we find $0.11(3)$~GeV$^{-4}$. If, for practical reasons, one insists in saturating with only two resonances, one would obtain $\Lambda_{a_1(1640)}=0.47(12)$~GeV.

With these numbers at hand, one could estimate missing isotriplet axial-vector-mesons contributions. Employing the results above for $B_{2S}^{a_1'}(0,0)$ and using the parametrization in \cref{eq:axialB2Sope} with $\Lambda_{a_1'}=1$~GeV,\footnote{Our estimate $\Lambda_{a_1'}=0.47(12)$~GeV is only meaningful once a parametrization for the form factor (that might involve different scales at high and low energies) is chosen.  Actually, assuming an universal axial decay constant $F_{a_1} \sim F_{a_1(1260)}$~\cite{Afonin:2004yb}, the OPE result in \cref{eq:OPEaxialTFF} implies, as discussed, $\Lambda_{a_1'} \simeq 1$~GeV also for this case. As such, we take $\Lambda_{a1'}\simeq 1$~GeV as a more realistic estimate.} we obtain $a_{\mu}= 2.5(^{+3.1}_{-1.9})\times 10^{-11}$ for $a_1(1640)$. The error arises from those of $B_{2S}^{a_1'}(0,0)$. We emphasize that this contribution should be thought of as an estimate for the full tower of missing states rather than a single $a_1'$ axial-vector meson, as it fully saturates the left-hand side of \cref{eq:SRO0}. While the lack of data implies large uncertainties, this illustrates the potential of this method once more data becomes available. As a comparison, the Regge and holographic models discussed in the following section predict for the remaining isotriplet contributions $a_{\mu}=3.67\times10^{-11}$ and $a_{\mu}=1.43\times10^{-11}$, respectively.

\section{Phenomenological models and contact contribution}\label{sec:simplemodel}

It is clear from our main discussion that an approach where transverse degrees of freedom satisfy the chiral limit relation in \cref{eq:opeCONS,eq:modelVVAano} is desirable, albeit this requires an infinite number of axial-vector mesons, as explained. While constructing such a model is a challenging task, in the following subsection we introduce a simple Regge model that meets the required criteria for symmetric kinematics. This illustrates our comments and allows to explore the role of subtracted and contact contributions. Finally, we also compare to the results of holographic models, that successfully reproduce \cref{eq:opeCONS} and support the outcome from our simplified model.

\subsection{A simple Regge-like model}\label{sec:Reggemodel}

\begin{figure}[tbp]
    \centering
    \includegraphics[width=0.45\textwidth]{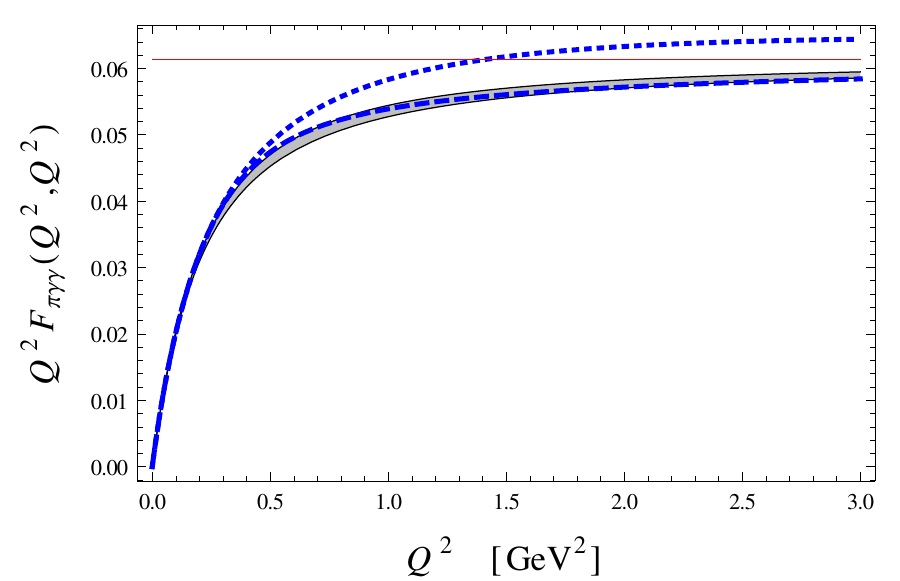}
    \caption{Comparison of this model prediction for the doubly-virtual $\pi^0$ TFF (gray band for $F_A\in(130,150)$~MeV) against that in Ref.~\cite{Masjuan:2017tvw}, whose doubly-virtual uncertainty is represented by the spread between the dotted and dashed lines (singly-virtual as well as systematic uncertainties are omitted). The solid red line shows the QCD prediction for its asymptotic limit.}
    \label{fig:pi0TFF}
\end{figure}

We construct a simplistic model with an infinite number of axial-vector mesons that reproduces the results described above, focusing on the symmetric form factor $B_{2S}$ that is known to play the main role in the HLbL. In particular, we take for the $n$-th axial vector meson excitation $A_n^a$ with flavor $a$ the following form factor
\begin{equation}
    B_{2S}^{A^a_n}(q_1^2,q_2^2) = 
    \frac{4F_A\operatorname{tr}(\mathcal{Q}^2\lambda^a)m_{A_n^a}}{[q_1^2 +q_2^2 -(M_a^2 +n\Lambda^2)]^2},
\end{equation}
for $n=0,1,2,...$, and assume that axial-vector mesons organize according to pure isotriplet, light-quark, and strange states [$a=\{3,q,s\}$ following the notation in \cref{eq:opeisospin}], with associated lowest-lying states $a_1(1260), f_1(1285), f_1(1420)$. Such construction ensures the OPE limit in \cref{eq:OPEaxialTFF} with the assumption that $F_A$ is a universal parameter for all of the resonances~\cite{Afonin:2004yb}. Restricting to the special case of symmetric spacelike $Q_1^2=Q_2^2$ kinematics, \cref{eq:modelVVAano} implies, in the chiral limit,
\begin{multline}
    \frac{N_c \operatorname{tr}(\mathcal{Q}^2 \lambda^a)}{4\pi^2}\left[ 1 -\tilde{F}_{P\gamma\gamma}(Q^2,Q^2)\right] = \\
   =  \lim_{Q_{1,2}^2\to Q^2}\sum_A \frac{F_A}{m_{A_n^a}}(Q_1^2 +Q_2^2)B_{2S}(Q_1^2,Q_2^2) +... \\
    = \lim_{Q_{1,2}^2\to Q^2} \frac{4F_A^2(Q_1^2 +Q_2^2)}{3\Lambda^4}\psi^{(1)}\left(\frac{M_a^2 +Q_1^2 +Q_2^2}{\Lambda^2}\right) +... \ ,
\end{multline}
with the ellipsis standing for antisymmetric form factors (irrelevant in this limit) and with $\psi^{(m)}(z)= \partial_{z}^{m+1}\ln\Gamma(z)$ the polygamma function. Taking the limit of large $Q_1^2=Q_2^2\equiv Q^2$, and using the result in \cref{eq:ptffOPE} for the pGB TFF, the equation above implies 
\begin{equation}
    1 -\frac{8\pi^2F_{P}^2}{3Q^2} = \left(\frac{4\pi F_A}{\Lambda \sqrt{N_c}}\right)^2\left[ 1+ \frac{\Lambda^2 -2M_a^2}{4Q^2}  \right]\,,
\end{equation}
up to order $\mathcal{O}(Q^{-4})$ corrections and independently of the axial-vector-mesons Regge trajectory. This requires $\Lambda=4\pi F_A N_c^{-1/2} \simeq 1$~GeV for $F_A=140$~MeV (c.f. footnote 11) as well as $M_a^2 = (8\pi^2/3)[F_A^2 +2F_P^2] \simeq 1$~GeV for isotriplet $a=3$. It also predicts $B_{2S}^{a_1}(0,0)=0.246~\textrm{GeV}^{-2}$, in agreement with our estimate from \cref{sec:axial}. Note also that, having fixed the parameters, we have a prediction for the pGB TFF. While for the $\eta$ and $\eta'$ mesons chiral and large-$N_c$ corrections are likely relevant to prevent such a straightforward connection, for the isotriplet channel one could expect a reasonable performance. Indeed, the prediction for the slope of the $\pi^0$ reads
\begin{equation*}
    b_{\pi} = \frac{3\psi^{(1)}\left(\frac{F_A^2 +2F_{\pi}^2}{2F_A^2}\right)}{16\pi^2 F_A^2} \sim 1.77(7)~\textrm{GeV}^{-2},
\end{equation*}
with the error from taking $F_A=140(10)$~MeV, in good agreement with the results in Refs.~\cite{Masjuan:2017tvw,Hoferichter:2018kwz}. In addition, we compare in \cref{fig:pi0TFF} the prediction for the doubly-virtual $\pi^0$ TFF (that is also insensitive to the contribution of axial-vector meson antisymmetric form factors) to the results in Ref.~\cite{Masjuan:2017tvw}, showing an excellent agreement. 

\begin{figure}[tbp]
    \centering
    \includegraphics[width=0.45\textwidth]{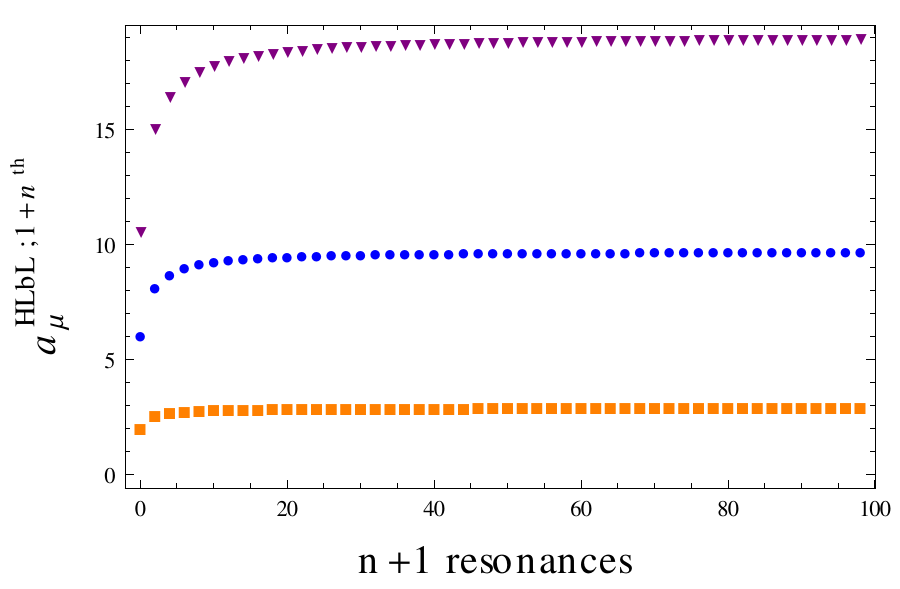}
    \caption{The contribution to $a_{\mu}^{\textrm{HLbL}}$ in units of $10^{-11}$ from the first $n+1$ $a_1$-, $f_1$-, and $f_1'$-like resonances as blue circles, purple triangles and orange squares, respectively.}
    \label{fig:ReggeSum}
\end{figure}

Next, we turn to the results of the model for the HLbL contribution to $a_{\mu}$. In the following, we will take the parameter $\Lambda=4\pi F_A N_c^{-1/2}$ in order to comply with the anomaly and $M_3^2 = (8\pi^2/3)[F_A^2 +2F_{\pi}^2]=980$~MeV, but choose $M_{q,s}$ to reproduce the experimental value for $B_{2S}^{f_1,f_1'}(0,0)$, that requires $M_{q(s)}=1.10(0.89)$~GeV---in agreement with the experimental ones. 
Finally, we take a Regge model for the axial-vector meson trajectories. In particular, we choose (Ref.~\cite{Masjuan:2012gc}):
\begin{align*}
    m_{a_1^n}^2 ={}& m_{a_1(1260)}^2 + n \mu^2_3\,, \\
    m_{f_1^n / f_1^{\prime n}}^2 ={}& m_{f_1(1285) / f_1(1420)}^2 + n \mu^2_{0},
\end{align*}
with $\mu^2_{3/0}=1.36/1.19~\textrm{GeV}^2$ from Ref.~\cite{Masjuan:2012gc}. Taking $F_A=140$~MeV, we find the results in \cref{tab:modelResults} for the sequential sum of the first $n+1$ resonances. In particular, we find a nice convergence considering the first 100 resonances, as it can be appreciated from \cref{fig:ReggeSum}. 

This model suggests a non-negligible contribution from heavy states that are necessary to comply with the OPE. Considering the one from all but the lowest-lying multiplet to be part of the short-distance constraints, this suggests $a_{\mu}^{\textrm{HLbL;SD}}=13\cdot10^{-11}$, in  line with the estimate in Ref.~\cite{Aoyama:2020ynm}. The total contribution agrees with holographic estimates, while that from excited states seems bigger in this model. We emphasize in this respect that we did not assess the error and that our model is incomplete, since it considers the symmetric form factors alone. Still, it reinforces the findings in \cite{Leutgeb:2019gbz,Cappiello:2019hwh} and serves as a simplistic illustration of our discussions. 

\begin{table}[tbp]
    \centering
    \begin{tabular}{c c c c c c c c} \hline
       n  & $0$ & $1$ & $5$ & $10$ & $40$ & $99$ & $(1-99)$ \\ \hline
       $a_1$  & $5.89$ & $7.35$ & $8.73$ & $9.12$ & $9.48$ & $9.56$ & $3.67$ \\ 
       $f_1$  & $10.52$ & $13.55$ & $16.84$ & $17.83$ & $18.77$ & $18.98$ & $8.46$\\
       $f_1'$  & $1.97$ & $2.35$ & $2.69$ & $2.77$ & $2.85$ & $2.87$ & $0.90$\\ \hline
       Total  & $18.38$ & $23.25$ & $28.26$ & $29.71$ & $31.10$ & $31.41$ & $13.03$\\       \hline
    \end{tabular}
    \caption{The contributions for the first $n+1$ resonances ($n=0$ is that of the lowest-lying state) to $a_{\mu}^{\textrm{HLbL}}$ for $a_1,f_1,f_1'$-like states in units of $10^{-11}$. The last column, $(1-99)$, shows the total contribution for $n=1$ to $n=99$ first states (e.g., the lowest-lying state has been subtracted).}
    \label{tab:modelResults}
\end{table}

\begin{table}[tbp]
    \centering
    \begin{tabular}{c c c c c c c c} \hline
       n  & $0$ & $1$ & $5$ & $10$ & $40$ & $99$ & $(1-99)$ \\ \hline
       Std  & $18.4$ & $23.3$ & $28.3$ & $29.7$ & $31.1$ & $31.4$ & $13.0$ \\ 
       Subt  & $-2.9$ & $-3.7$ & $-4.5$ & $-4.8$ & $-5.0$ & $-5.1$ & $-2.2$\\
       Cont  & $21.3$ & $27.0$ & $32.9$ & $34.6$ & $36.2$ & $36.6$ & $15.3$\\  \hline
    \end{tabular}
    \caption{Axial-vector-mesons contributions to $a_{\mu}^{\textrm{HLbL}}$ in units of $10^{-11}$. We show the decomposition into Contact and Subtracted parts, which add up to the Standard result. We find that, having the contact term and the first subtracted contribution from the lowest-lying axial meson, the remaining part (last column in Subt row) column is negligible.}
    \label{tab:modelDecomp}
\end{table}

Finally, in \cref{tab:modelDecomp}, we decompose the standard result (see \cref{tab:modelResults}) into the sum of subtracted and contact contributions. As suggested in \cref{sec:axialsHLBL}, the contact term provides the dominant contribution for the whole tower of axial-vector meson contributions. Further, incorporating on top the subtracted contributions from the lowest-lying axial-vector mesons, one obtains an excellent estimate of the whole result. To further speculate on the possibility that such might be a general feature, in the subsection below, we pursue a similar study employing the (more sophisticated) holographic model from Ref.~\cite{Leutgeb:2019gbz}, that also complies with the anomaly requirements.

\subsection{A holographic model}

In what follows, we use the holographic results (HW2 model) used in Ref.~\cite{Leutgeb:2019gbz} to compute the axial-vector meson contributions to $a_{\mu}$, decomposing their contributions into the subtracted and contact part. Taking their form factors, we reproduce their results for individual axial-vector-mesons contributions in \cref{tab:Holo}, ``Std'' row (compare to Table~3 in Ref.~\cite{Leutgeb:2019gbz}). In addition, \cref{tab:Holo} displays the splitting into contact (``Cont'') and subtracted (``Subt'') parts. In each of these rows, the first subrow collects the result for each individual resonance and the second one the accumulated contribution up to and including the given resonance in each column.  As expected, the contact term dominates the complete result for each axial-vector meson, and thereby for the whole tower. In addition, the sum of the contact term and the subtracted contribution from the lowest-lying axial-vector meson states provides an excellent estimate of the whole contribution, even if the relative weight of individual contributions differs from our model. Moreover, the full result is pretty similar in both models. A plausible explanation is that the contact term might have little model-dependence provided it successfully reproduces the pGB TFF, as we find in \cref{fig:pi0TFF}.

\begin{table}[tbp]
\begin{tabular}{cccccc} \hline
            & $j=1$      & $j \leq 2$  &  $j \leq 3$ & $j \leq 4$  & $j \leq 5$ \\ \hline
 Std   & $23.0(7)$  & $3.2(2)$    &  $1.17(1)$  & $0.50(4)$   & $0.27(2)$ \\  
            & $23.0(7)$  & $26.2(7)$   &  $27.4(7)$  & $27.9(7)$   & $28.1(7)$ \\  
 Subt & $-4.35(10)$ & $-0.70(5)$  & $-0.24(1)$  & $-0.11(1)$  & $-0.06(0) $ \\ 
            & $-4.35(10)$ & $-5.05(10)$ & $-5.29(10)$ & $-5.40(10)$ & $-5.46(10)$ \\ 
 Cont  & $27.4(3)$  & $3.92(8)$    &  $1.40(3)$  & $0.61(4)$   & $0.31(1)$ \\ 
            & $27.4(3)$  & $31.3(3)$   &  $32.7(3)$  & $33.3(3)$   & $33.6(3)$ \\ \hline
\end{tabular}
\caption{\label{tab:Holo}Axial-vector-mesons contributions to $a_{\mu}^{\textrm{HLbL}}$ in units of $10^{-11}$ according to the HW2 model from Ref.~\cite{Leutgeb:2019gbz} split into standard (Std), subtracted (Subt) and contact (Cont) terms. Each column displays the contribution from the $j$-th multiplet and the sequential summation immediately below. The errors are from the numerical integration.}
\end{table}

The lesson drawn here is the following: suppose that only the lowest-lying axial-vector meson is accessible experimentally, but not the rest; suppose we have a phenomenological model which satisfies the anomaly constraint \cref{eq:modelVVAano}. You want to determine $a_{\mu}^{\textrm{HLbL}}$ with best precision. 
In order to achieve that, you should take the decomposition
in \cref{eq:axialPROP} and compute the whole contact term contribution from the model. As long as the model reproduces \cref{eq:modelVVAano}, such contribution should have little model dependence. Add on top the subtracted contribution from the experimentally-known lowest-lying states. The remaining part would be subleading, and beyond our current needs in evaluating $a_{\mu}^{\textrm{HLbL}}$. Still, one could use the model estimate for the remaining subtracted part, where one would expect a moderate model dependence.

\section{Comments and outlook} \label{sec:Concl}

In this work, we have revisited the OPE constraint derived in Ref.~\cite{Melnikov:2003xd} in the context of the anomalous magnetic moment of the muon, that relates the HLbL Green's function to the $\langle VVA \rangle$ one. In particular, we paid special attention to the role of the pseudoscalar poles and transverse contributions. As a result, we find unexpected relations among them, which are required to fulfill the axial anomaly and that have not received attention in the past. The consequences are important and allow, among others, to understand the  discrepancies among different results in the literature. In particular, regarding the role of axial-vector mesons, we have been able to overcome the problem first outlined in Ref.~\cite{Roig:2019reh}, that concerns the basis ambiguities when describing the axial-vector mesons TFFs---a key ingredient when computing their contribution to $a_{\mu}$. To do so, the anomaly played a central role, since it fixes the (potentially ambiguous) nonpole terms. This was better illustrated by splitting these contributions into a contact and a subtracted part, offering suggestive possibilities in the future to incorporate OPE constraints and also clarifying the precise meaning of the results in RChT. The receipt here proposed is exemplified with the help of two different models, both of them including a tower of axial-meson contributions.

In the future, it might be interesting to study further implications related to well-known high-energy behavior of the $\langle VVA \rangle$ Green's function. In addition, it might be interesting to discuss finite mass corrections that were left aside in this work, and are not related to the anomaly, but to the $\langle VVP \rangle$ Green's function (and the $U_A(1)$ gluon anomaly for the singlet current). While these should not affect the discussion concerning the role of axial-vector mesons, it might help constraining the effect of heavy pseudoscalar resonances in the HLbL, that are present when considering finite quark masses. This is especially relevant concerning the octet and singlet current, and represents an important and challenging point as outlined in Ref.~\cite{Aoyama:2020ynm}.

\section*{Acknowledgments}
The authors are indebted to R.~Escribano and S.~Peris for useful discussions and for encouraging at early stages of this work. P. S. acknowledges M.~Knecht for enlightening conversations motivating this work and for useful comments on this manuscript. We also acknowledge M.~Hoferichter for insightful observations on this manuscript. P. M and P. S. received financial support from European Regional Development Funds under the Spanish Ministry of Science, Innovation and Universities (project  FPA2017-86989-P), from the Agency for Management of  University  and Research  Grants  of  the Government of Catalonia (project SGR 1069) and  by  the  COFUND  program  of  the  Marie Sklodowska-Curie actions under the framework program Horizon  2020  of  the  European  Commission,  Grant No. 754510 (EU, H2020-MSCA-COFUND2016). P. R. has been partially funded  through the project 250628 (Ciencia B\'asica, Conacyt), Fondo SEP-Cinvestav 2018 (project number 142) and C\'atedras Marcos Moshinsky 2020 (Fundaci\'on Marcos Moshinsky).  The research of P. S. is also supported by Ministerio de Industria, Econom\'ia y Competitividad under the grant SEV-2016-0588.

\bibliographystyle{apsrev4-1}
\bibliography{bibfile}% Produces the bibliography via BibTeX.

\end{document}